\documentclass[aps,prl,twocolumn,groupedaddress,amsmath,amssymb,10pt]{revtex4-1}
\usepackage{graphicx,xcolor}  
\usepackage{dcolumn}   
\usepackage{bm}        
\usepackage{verbatim}   
\usepackage{soul} 

\newcommand{\rom}[1]{%
  \textup{\uppercase\expandafter{\romannumeral#1}}%
}

\newcommand{\hc}{}

\begin{document}

\title{Invariant Solution underlying Oblique Stripe Patterns in Plane Couette Flow}
\author{Florian Reetz}
\affiliation{
Emergent Complexity in Physical Systems Laboratory (ECPS),\\
\'Ecole Polytechnique F\'ed\'erale de Lausanne, CH 1015 Lausanne, Switzerland
   }
\author{Tobias Kreilos}
\affiliation{
Emergent Complexity in Physical Systems Laboratory (ECPS),\\
\'Ecole Polytechnique F\'ed\'erale de Lausanne, CH 1015 Lausanne, Switzerland
   }
   \author{Tobias M. Schneider}
\affiliation{
Emergent Complexity in Physical Systems Laboratory (ECPS),\\
\'Ecole Polytechnique F\'ed\'erale de Lausanne, CH 1015 Lausanne, Switzerland
   }
\date{\today}
\begin{abstract}
\noindent 
When subcritical shear flows transition to turbulence, laminar and turbulent flow often coexists in space, giving rise to turbulent-laminar patterns. Most prominent are regular stripe patterns with large-scale periodicity and oblique orientation. \hc{Oblique stripes are a robust phenomenon, observed in experiments and flow simulations, yet their origin remains unclear.} We demonstrate the existence of an invariant equilibrium solution of the fully nonlinear 3D Navier-Stokes equations that resembles the oblique pattern of turbulent-laminar stripes in plane Couette flow. We uncover the origin of the stripe equilibrium and show how it emerges from the well-studied Nagata equilibrium via two successive symmetry-breaking bifurcations. 
\end{abstract}

\maketitle

\noindent The complex laminar-turbulent transition in wall-bounded shear flows is one of the least understood phenomena in fluid mechanics. In the simple geometry of plane Couette flow (PCF), the flow in a gap between two parallel plates moving in opposite directions, the transitional flow
spontaneously breaks the translational symmetries in both the streamwise and the spanwise direction causing regions of turbulent and laminar flow to coexist in space \cite{Lundbladh1991,Tillmark1992,Daviaud1992a,Lemoult2016,Couliou2017}. Remarkably, the flow may further self-organize into a regular pattern of alternating turbulent and laminar stripes \cite{Prigent2002,Barkley2005,Barkley2007,Duguet2010,Tuckerman2011,Philip2011,Ishida2017} also observed in Taylor-Couette \cite{Coles1965,Andereck1986,Hegseth1989,Prigent2002,Meseguer2009,Dong2009} and channel flow \cite{Tsukahara2005,Hashimoto2009,Aida2014,Tuckerman2014a,Xiong2015}. The wavelength of these stripes or bands is much larger than the gap size, the only characteristic scale of the system, and they are oblique with respect to the preferred streamwise direction. \hc{Consequently, both the large-scale wavelength and the oblique orientation of turbulent-laminar stripes, must directly follow from the flow dynamics captured by the governing Navier-Stokes equations. Experiments and numerical flow simulations reliably generate stripe patterns but a theory explaining the origin of the pattern characteristics is still missing. This is related to the Navier-Stokes equations being highly nonlinear partial differential equations, whose theoretical analysis remains challenging. }\\
\indent It was the early observation that an oblique turbulent-laminar pattern can be the preferred solution of the Navier-Stokes equations that motivated R. Feynman to stress the lack of ``mathematical power [of his time] to analyze [the Navier-Stokes equations] except for very small Reynolds numbers'' \cite{Feynman1964}. \hc{Recent advances in numerical methods make it possible to not only simulate flows but to construct exact equilibria, traveling waves and periodic orbits of the fully nonlinear 3D Navier-Stokes equations. These dynamically unstable \emph{exact invariant solutions} lead to a description of turbulence as a chaotic walk among invariant solutions which together with their entangled stable and unstable manifolds support the turbulent dynamics \cite{Lanford1982,Gibson2008}. Exact invariant solutions are thus `building blocks' which resemble characteristic flow structures that are observed in flow simulations and experiments, when the dynamics transiently visits the invariant solution. A theoretical explanation of oblique stripe patterns within this \emph{dynamical systems description} requires the yet unsuccessful identification of exact invariant solutions resembling the detailed spatial structure of turbulent-laminar stripes, including their oblique orientation and large-scale periodicity.}\\
\indent Nagata discovered the first invariant solution of PCF \cite{Nagata1990}. Like most invariant solutions of PCF found since then \cite{Gibson2009}, this so-called Nagata equilibrium is periodic in the streamwise and spanwise directions. Spatially periodic solutions of this type do not capture the coexistence of turbulent and laminar flow and consequently cannot underly oblique stripes. Spanwise localized invariant solutions \cite{Schneider2010,Gibson2014} and doubly localized invariant solutions in extended periodic domains \cite{Brand2014}, show nonlinear flow structures coexisting with laminar flow but no known solution captures oblique orientation or suggests a pattern wavelength matching oblique stripe patterns.\\
\begin{figure}[b!]
\includegraphics[width=0.99\linewidth]{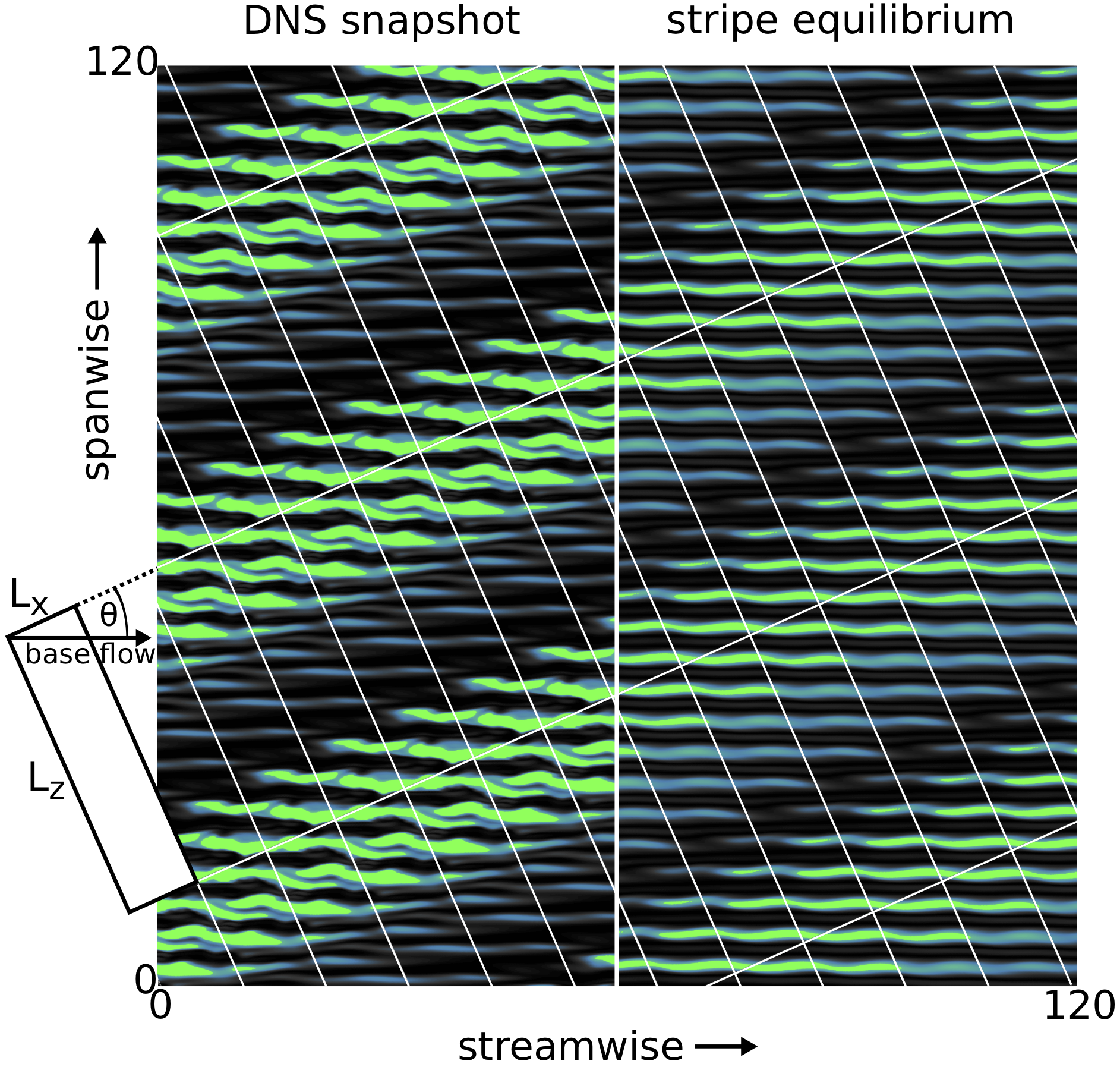}
\caption{\label{fig:comparison} Oblique turbulent-laminar stripes observed at $Re=350$ in DNS (left) and the underlying invariant equilibrium solution (right). Following Ref.~\citep{Barkley2005} a tilted periodic domain outlined on the left side with $(\theta, L_x, L_z)=(24^{\circ},10,40)$ is used for computations. The contours are turbulent kinetic energy in a plane at 3/4 of the gap height and saturate at $u^2=0.25$ (green).}
\end{figure}
\indent We present a fully nonlinear equilibrium solution of PCF (right half of Fig.~\ref{fig:comparison}), resembling the oblique stripe pattern observed in direct numerical simulations (left half in Fig.~\ref{fig:comparison}). \hc{Parametric continuation demonstrates that this stripe equilibrium} is connected to the well-studied Nagata equilibrium via two successive symmetry-breaking bifurcations. The bifurcation scenario suggests a spatial phase shift mechanism to underlie the formation of turbulent-laminar patterns in extended shear flows.\\
\indent \emph{Numerical simulations.} For direct numerical simulations (DNS) of oblique stripe patterns in PCF we use a parallelized version of the pseudo-spectral code \emph{Channelflow} \cite{Gibson2008}. The numerical domain is periodic in two perpendicular dimensions along the plates ($x$ and $z$) with periods of $(L_x,L_z)=(10,40)$ in half-gap units. Inversion symmetry with respect to the domain center is enforced. The relative plate velocity and the associated base flow are tilted against the periodic domain dimensions at an angle of $\theta=24^{\circ}$ following Barkley \& Tuckerman \citep{Barkley2005}. At Reynolds number $Re=Uh/\nu=350$, with the relative wall velocity $2U$, gap height $2h$ and kinematic viscosity $\nu$ the flow organizes into self-sustained turbulent-laminar stripes, as shown in  Fig.~\ref{fig:comparison} (left), where we periodically repeat the computational domain to highlight the large-scale structure of the pattern. \\
\indent \emph{Stripe equilibrium.} An invariant equilibrium solution capturing the stripes was found by introducing a large-scale amplitude modulation to a known invariant solution using a suitable window function, similar to Ref.~\cite{Gibson2014}. Specifically, the Nagata equilibrium was periodically extended in the spanwise direction for $n=9$ periods, then sheared to align the velocity streaks with the base flow in the tilted domain and finally  multiplied with a scalar window function equal to a scaled mean field of turbulent kinetic energy of the oblique stripe pattern from several DNS runs. Using the constructed velocity field as initial guess Newton iteration converges to the stripe equilibrium (Fig.~\ref{fig:comparison}, right).\\
\indent The stripe equilibrium shares the small-scale wavy modulation with the Nagata equilibrium but also shows the large-scale oblique amplitude modulation of the turbulent-laminar stripe pattern. The amplitude modulation between the high-amplitude \emph{turbulent region} and the low-amplitude \emph{laminar region} of the equilibrium on average follows a sinusoidal profile closely resembling the pattern mean flow found in DNS at identical boundary conditions \citep{Barkley2007}. The stripe equilibrium moreover captures detailed features of the turbulent-laminar interfaces. A base flow directed into a turbulent region leads to a sharper 'upstream' interface than a base flow directed out of a turbulent region at a 'downstream' interface.
The direction of the base flow is reversed for $y \to -y$. An upstream interface in the upper half thus corresponds to a downstream interface in the lower one. This gives rise to so-called \emph{overhang regions} \cite{Lundbladh1991, Duguet2013} and an asymmetry between the left and right interface in Fig.~\ref{fig:comparison} where turbulent kinetic energy is visualized at $y=0.5$ above the midplane. Finally, the stripe equilibrium is symmetric under inversion $\sigma_i[u,v,w](x,y,z) = [-u, -v,-w](-x,-y,-z)$, a symmetry also found for the mean flow of stripe patterns  \citep{Barkley2007}. The sinusoidal amplitude modulation, the captured overhang regions and the inversion symmetry, all characteristic of the pattern's mean flow, together with the visual comparison in Fig.~\ref{fig:comparison} show that the stripe equilibrium has the spatial features of the oblique stripe pattern. We have thus identified a first exact invariant solution underlying the oblique turbulent-laminar patterns. \\
\begin{figure}
\includegraphics[width=0.95\linewidth]{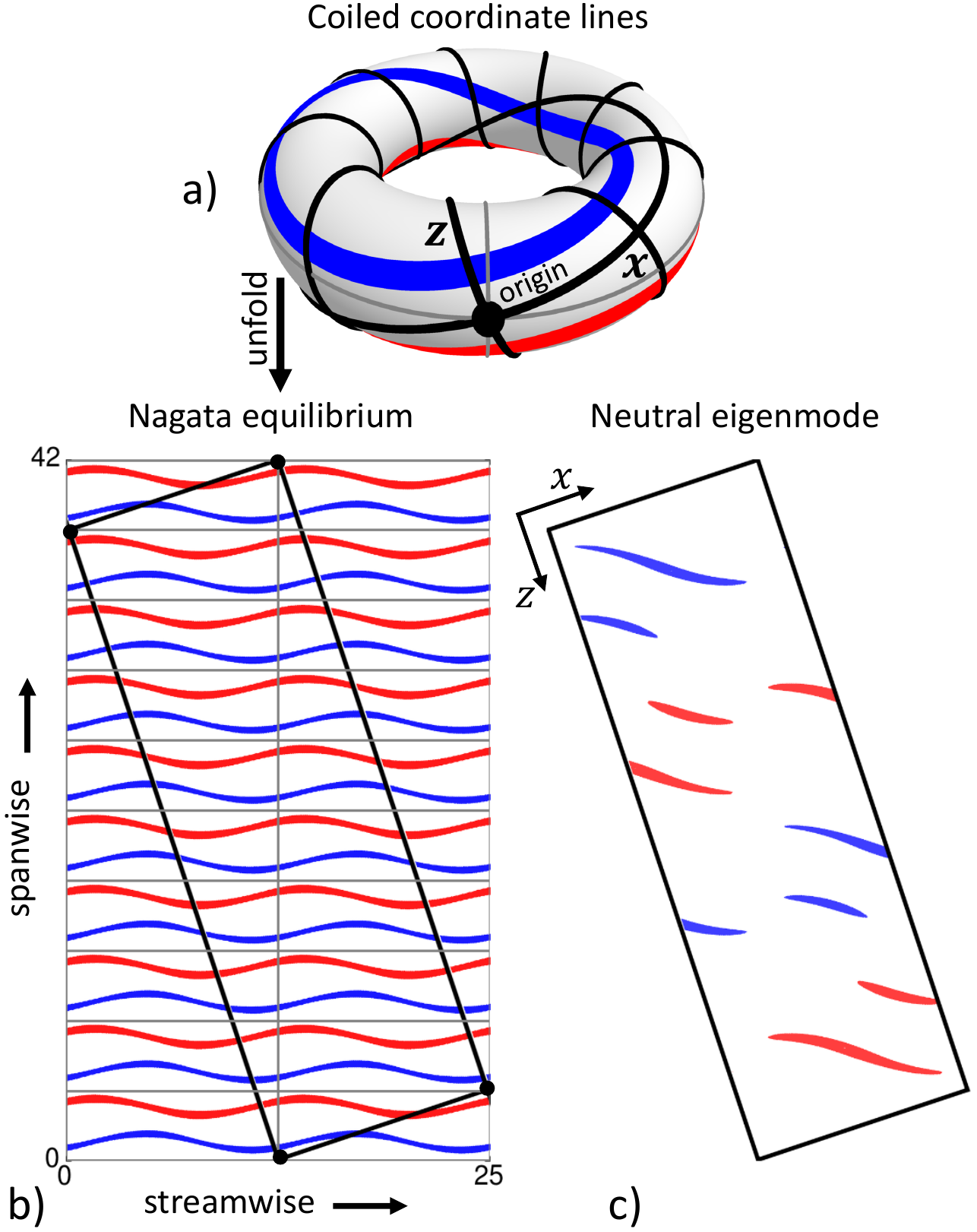}
\caption{\label{fig:nbcwinstab} 
Oblique modulational instability of the Nagata equilibrium:
(a) Torus representing a streamwise-spanwise periodic domain. If tilted rectangular coordinate lines $x$ and $z$ (black) close on themselves, all solutions on the torus also respect the periodicity of a domain spanned by those lines. 
Specifically, the Nagata equilibrium with streamwise-spanwise periodicity  $(\lambda_{st},\lambda_{sp})=(12.65,4.22)$ (grey lines) is also periodic with respect to the tilted domain (black) with  $(\theta,L_x,L_z)=(18.4^{\circ},40/3,40)$ (b). In this tilted domain, a bifurcation with neutral eigenmode (c) can be detected. This bifurcation introduces oblique long-wavelength amplitude modulations on the Nagata equilibrium. 
Red (blue) contours represent positive (negative) downstream velocity in the midplane. }
\end{figure}
\indent \emph{Nagata equilibrium.} At small scales the stripe equilibrium reflects the wavy streak structure of the spatially periodic Nagata equilibrium. This suggests the stripe equilibrium to emerge from the Nagata equilibrium in a bifurcation creating oblique long-wavelength modulations.
To identify this pattern-forming bifurcation numerically, the Nagata equilibrium needs to 'fit' in an extended tilted periodic domain aligned with the wave-vector of the neutral mode creating the oblique long-wavelength modulation. The Nagata equilibrium indeed not only satisfies the stream- and spanwise periodic boundary conditions of the commonly studied minimal flow domain but may also be periodic with respect to selected larger tilted domains. The symmetry group of the Nagata equilibrium, including all combined discrete translations over streamwise-spanwise periods $(\lambda_{st},\lambda_{sp})$, intersects with the group of translations of a tilted rectangular domain, periodic at $(L_x,L_z)$, if
\begin{align}
\label{eq:domain_condition}
L_x=\frac{k\,\lambda_{st}}{\cos\theta}=\frac{l\,\lambda_{sp}}{\sin\theta} \ ,\quad
L_z=\frac{m\,\lambda_{st}}{\sin\theta}=\frac{n\,\lambda_{sp}}{\cos\theta}
\end{align}
is satisfied for $(k,l,m,n)\,\in\,\mathbb{N}$ and $0<\theta<\pi/2$. Geometrically, condition (\ref{eq:domain_condition}) describes how the $x$-$z$ coordinate lines of the tilted domain wind on a torus defined by the streamwise-spanwise periodic minimal domain. The condition is satisfied if the coordinate lines are closed curves (Fig.~\ref{fig:nbcwinstab}a). 
For the domain $(\theta, L_x, L_z)=(24^{\circ},10,40)$ considered so far, the geometric condition  (\ref{eq:domain_condition}) implies wavelengths 
$(\lambda_{st},\lambda_{sp})=(1.02,4.06)$ at which the Nagata equilibrium does not exist. Keeping $L_z=40$ and choosing winding numbers $(k,l,m,n)=(1,1,1,9)$ however leads to a slightly modified domain $(\theta, L_x, L_z)=(18.4^{\circ},40/3,40)$  in which the Nagata equilibrium with $(\lambda_{st},\lambda_{sp})=(12.65,4.22)$ exists, as displayed in Fig.~\ref{fig:nbcwinstab}b. On the lower branch of the Nagata equilibrium close to the saddle-node, there is a pitchfork bifurcation at $Re_{\rom{1}}=164$. Its neutrally stable long-wavelength eigenmode is plotted in Fig.~\ref{fig:nbcwinstab}c. This is the initial pattern-forming bifurcation creating oblique amplitude modulations on the Nagata equilibrium.\\
\indent \emph{Bifurcation scenario.} Using parametric continuation we follow both the periodic Nagata equilibrium (named $\mathcal{A}$ hereafter) and the emerging modulated equilibrium solution ($\mathcal{B}$) from its primary bifurcation point at $(Re,\theta,L_x)_{\rom{1}}=(164,18.4^{\circ},40/3)$ to the parameters $(Re,\theta,L_x)_{\mathcal{C}}=(350,24^{\circ},10)$ of the stripe equilibrium ($\mathcal{C}$). In the three dimensional parameter space we choose a continuation path parametrized by tilt angle $\theta$ with the Reynolds number linear in $\theta$, such that $ Re(\theta)=\left( Re_{\rom{1}} (\theta_{\mathcal{C}}-\theta) + Re_{\mathcal{C}} (\theta-\theta_{\rom{1}})\right)/(\theta_{\mathcal{C}}-\theta_{\rom{1}}) $ and domain length $L_x(\theta)=L_z/(n\,\tan(\theta))$ for $n=9$ and constant domain width of $L_z=40$. The resulting bifurcation diagram demonstrates that the Nagata equilibrium $\mathcal{A}$, is connected to the stripe equilibrium $\mathcal{C}$ (Fig.~\ref{fig:bifplot}).\\
\begin{figure}
\includegraphics[width=0.99\linewidth]{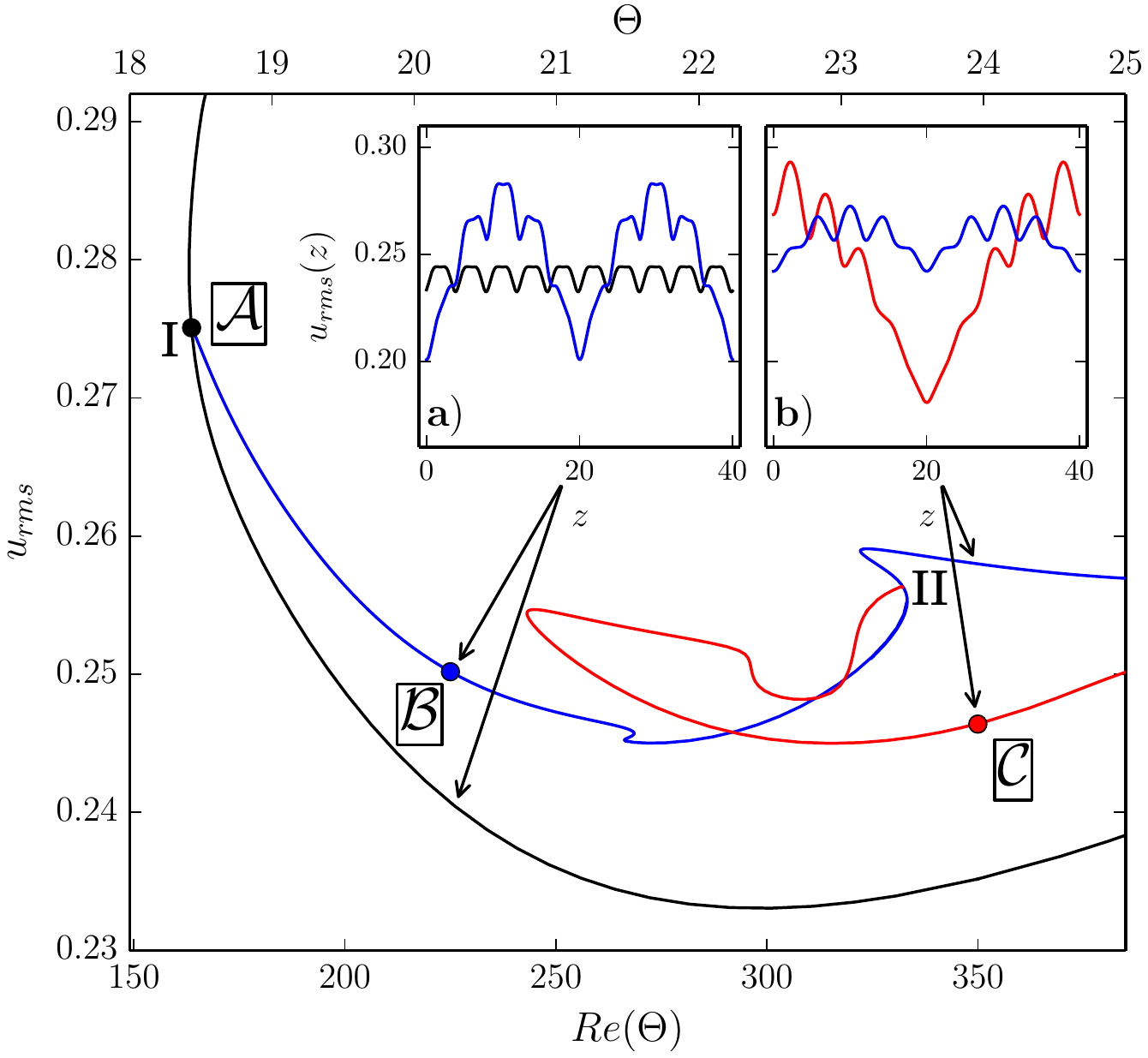}
\caption{\label{fig:bifplot} Pattern forming bifurcation sequence from the small-scale periodic Nagata equilibrium $\mathcal{A}$ (Fig.~\ref{fig:nbcwinstab}) to the large-scale modulated stripe equilibrium $\mathcal{C}$ (Fig.~\ref{fig:comparison}). The solution branches indicate domain averaged velocity norm $u_{\text{rms}}$ over linearly coupled bifurcation parameters $\theta$ and $Re(\theta)$ (see text). A primary pattern-forming bifurcation on $\mathcal{A}$ at $(Re,\theta)_{\rom{1}}=(164,18.4^{\circ})$ creates equilibrium solution $\mathcal{B}$ having a double-pulse profile of $x$-$y$ averaged $u_{\text{rms}}(z)$ (inset a) and bifurcating in a secondary pattern-forming bifurcation at $(Re,\theta)_{\rom{2}}=(332,23.4^{\circ})$ to the single-pulse solution branch of equilibrium $\mathcal{C}$ (inset b). Points mark the invariant solutions shown in Fig.~\ref{fig:phase}.}
\end{figure}
\indent The primary bifurcation is of pitchfork type, subcritical, forward and breaks the streamwise-spanwise translation symmetry of $\mathcal{A}$. Along the bifurcating branch of $\mathcal{B}$ significant amplitude modulations of the small scale periodic signal form with period $L_z/2$ along $z$, as indicated by the double-pulse profile of $u_{\text{rms}}(z)$ at $Re=225$ in Fig.~\ref{fig:bifplot}a. The modulation period reflects a discrete translation symmetry $\sigma_{\mathcal{B}}$ over half the domain diagonal, $\sigma_{\mathcal{B}}[u,v,w](x,y,z) = [u,v,w](x+L_x/2,y,z+L_z/2)$. Equilibrium $\mathcal{B}$ inherits this symmetry from $\mathcal{A}$ because $\sigma_{\mathcal{B}}$  is not broken by the neutral mode of the primary bifurcation (Fig.~\ref{fig:nbcwinstab}c). \\
\indent A secondary pattern-forming bifurcation occurs at $(Re,\theta,L_x)_{\rom{2}}=(332,23.4^{\circ},10.3)$ along solution branch $\mathcal{B}$ (blue line in Fig.~\ref{fig:bifplot}). This subcritical pitchfork bifurcation breaks the translation symmetry $\sigma_{\mathcal{B}}$. The spatial period of the modulation is doubled giving rise to the single-pulse equilibrium $\mathcal{C}$. The amplitude profiles of single- and double-pulse equilibrium show that the single-pulse with period $L_z=40$ has large modulations at $Re=350$ and $\theta=24^{\circ}$, while the modulations in the double-pulse equilibrium are reduced (Fig.~\ref{fig:bifplot}b). This agrees with the observations that stripes tend to have pattern wavelengths $\lambda$ in the range of $40 \leq \lambda\leq 60$ at $Re$ around 350 \cite{Prigent2002,Tuckerman2011}. Solution branch $\mathcal{C}$ (red line in Fig.~\ref{fig:bifplot}) reaches $Re_{\mathcal{C}} = 350$  after undergoing an additional saddle-node bifurcation at $(Re,\theta,L_x)=(243,20.8^{\circ},11.7)$. In summary, two  bifurcations successively break discrete translation symmetries of the Nagata equilibrium to create the stripe equilibrium solution. \\
\indent Small-scale velocity streaks carry a wavy modulation which has a downstream phase that is clearly evident when plotting the downstream vorticity at the midplane. We illustrate the relative phase by a line connecting vorticity maxima or minima in the spanwise direction (red lines in Fig.~\ref{fig:phase}). A straight and strictly spanwise oriented line indicates the identical downstream phase of all streaks of the (spanwise periodic) Nagata equilibrium $\mathcal{A}$.
The primary pattern forming bifurcation from $\mathcal{A}$ to $\mathcal{B}$ introduces local phase shifts which dislocate vorticity extrema away from a straight alignment. The dislocations introduced in $\mathcal{B}$ are symmetric with respect to half-domain translations $\sigma_{\mathcal{B}}$ and centered at $z=0$ and $z=20$. 
For the stripe equilibrium $\mathcal{C}$ formed from $\mathcal{B}$ in the second pattern forming bifurcation, the topology of the lines is preserved but they are geometrically deformed into sigmoidal structures. \\
\indent Locally, in the \emph{turbulent region} of equilibrium $\mathcal{C}$ this deformation implies uniform phase shifts indicated by a line skewed at $\theta=24^{\circ}$ against the spanwise direction and oriented exactly along the pattern wave vector. Such skewed relative phases have previously been observed within localized variants of the Nagata equilibrium \cite{Gibson2016}.
The \emph{laminar region}, in contrast, shows alternating phase shifts indicated by lines skewed at $\theta=\pm 24^{\circ}$ against the spanwise direction. On average over $x$, the dimension along which the stripe pattern is statistically homogeneous \citep{Barkley2007}, the phase shift in the \emph{laminar region} is zero.  The \emph{laminar} region thus shows phase relations resembling the Nagata equilibrium while the wavy streaks are skewed in the \emph{turbulent region} with phase relations aligned in the direction of the oblique stripe pattern. Consequently, phase relations and amplitude modulations are strongly correlated. This suggests a form of effective 'elastic response' of Nagata-type invariant solutions where deforming the streaks away from their preferred alignment requires higher energy and increases the amplitude. There appears to be a maximal deformation that a Nagata-type invariant solution can sustain \cite{Gibson2016}. Since an invariant stripe solution of $\mathcal{C}$-type requires wavy streaks in the \emph{turbulent region} to be skewed at the angle of the oblique pattern, the largest possible deformation of the Nagata equilibrium may thus control the angle at which oblique stripe patterns can exist.\\
\indent \hc{Experimental and numerical observations} of self-organized oblique turbulent-laminar stripes in wall-bounded extended shear flows suggest the existence of exact invariant solutions underlying these patterns. We present the first such invariant solution of the fully nonlinear 3D Navier-Stokes equations in plane Couette flow that captures the detailed spatial structure of oblique stripe patterns. The stripe equilibrium emerges from the Nagata equilibrium via a sequence of two pattern-forming bifurcations with long-wavelength oblique neutral modes. Spatial amplitude modulations in the stripe equilibrium are correlated with local phase shifts between velocity streaks suggesting phase modulations to drive the amplitude modulations on large scales which give rise to oblique turbulent-laminar patterns. 
\begin{figure}
\includegraphics[width=0.99\linewidth]{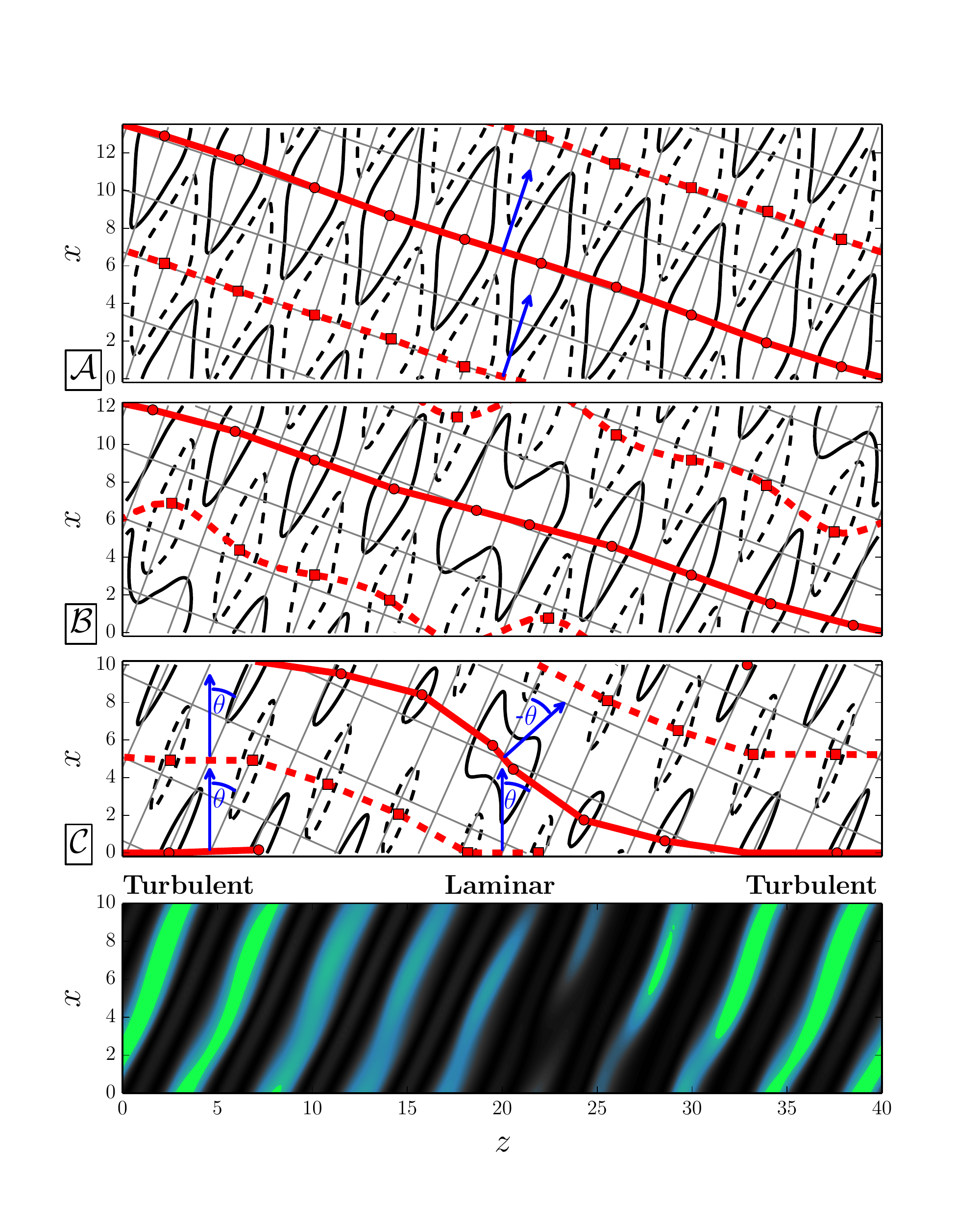}
\caption{\label{fig:phase} 
Downstream vorticity $\omega_{ds}=\vec{\nabla}\times\vec{u}\ (\cos(\theta)\vec{e}_x+\sin(\theta)\vec{e}_z)$ at the midplane (black solid/dashed contours at $\omega_{ds}=\pm0.12$) encoding downstream phase information of wavy streak modulations for equilibria $\mathcal{A}$, $\mathcal{B}$ and $\mathcal{C}$ along the bifurcation sequence (points in Fig.~\ref{fig:bifplot}). Red solid (dashed) lines guide the eye along vorticity maxima (minima). Lines skewed against the streamwise-spanwise directions (grey orthogonal grid) indicate phase shifts between neighboring streaks with blue arrows highlighting the angle $\theta$ of misalignment relative to the streamwise direction. 
In the Nagata equilibrium $\mathcal{A}$ straight red lines in the spanwise direction indicate zero phase shifts. For the stripe equilibrium $\mathcal{C}$ a sigmoidal red line encodes phase modulations: A uniform phase shift skews the wavy streaks  in the turbulent region while the average phase shift remains zero in the laminar region. These phase modulations correlate with amplitude modulations and create the oblique pattern of turbulent-laminar stripes (bottom panel, see also Fig.~\ref{fig:comparison}).}
\end{figure}
\end{document}